\documentclass[twocolumn,twocolappendix]{aastex631}

\usepackage{amsthm,amsmath}
\newtheoremstyle{newplain}% name
{}% space above
{}% space below
{\it}% body font
{}% indent amount
{\bfseries}% theorem head font
{.}% punctuation after theorem head
{5pt}% space after theorem head (default: 5pt)
{\thmname{#1}\hspace{5pt}\thmnumber{#2}\thmnote{\hspace{2pt}[{\normalfont #3}]}}% theorem head spec
\theoremstyle{newplain}
\usepackage{cases}
\usepackage{amsfonts}
\usepackage{mathrsfs,mathtools}
\usepackage{physics,amssymb}
\usepackage{braket}
\usepackage{savesym}
\savesymbol{tablenum}
\usepackage{siunitx}
\usepackage{acronym}
\usepackage{soul}

\newcommand{\calM}{\mathcal{M}}
\newcommand{\bfw}{\mathbf{w}}
\newcommand{\bfx}{\mathbf{x}}

\newcommand{\bfz}{\mathbf{z}}

\usepackage{graphicx}
\hypersetup{
colorlinks=true
,urlcolor=DARKBLUE
,anchorcolor=DARKBLUE
,citecolor=DARKBLUE
,filecolor=DARKBLUE
,linkcolor=DARKBLUE
,menucolor=DARKBLUE
,linktocpage=true
,pdfproducer=medialab
,pdfa=true
}
\usepackage{xcolor}
\usepackage{comment}
\newcommand{\red}[1]{\textcolor{red}{#1}}

\newcommand{\mr}[1]{\mathrm{#1}}
\usepackage{aas_macros}

\DeclareMathOperator{\erfc}{erfc}

\newcommand{\ee}{\mathrm{e}}
\newcommand{\diag}{\mathrm{diag}}

\newcommand{\uth}{\mathrm{th}}
\newcommand{\pk}{\mathrm{pk}}
\newcommand{\PBH}{\mathrm{PBH}}
\newcommand{\DM}{\mathrm{DM}}
\newcommand{\eff}{\mathrm{eff}}

\newcommand{\calC}{\mathcal{C}}

\newcommand{\uG}{\mathrm{G}}

\newcommand{\bfk}{\mathbf{k}}

\newcommand{\um}{\mathrm{m}}

\newcommand{\calO}{\mathcal{O}}

\newcommand{\calP}{\mathcal{P}}

\newcommand{\beae}[1]{\begin{equation}\begin{aligned} #1 \end{aligned}\end{equation}}

\newcommand{\bae}[1]{\begin{align} #1 \end{align}}
\newcommand{\bce}[1]{\begin{cases} #1 \end{cases}}

\newcommand{\bme}[1]{\begin{multline} #1 \end{multline}}

\newcommand{\relBigm}[1]{\mathrel{}\Bigm#1\mathrel{}}

\definecolor{MONZA}{HTML}{CF000F}
\definecolor{DARKBLUE}{HTML}{00008b}
\definecolor{DARKMAGENTA}{HTML}{8b008b}
\definecolor{DARKORANGE}{HTML}{ff8c00}

\newcommand{\Blue}[1]{\textcolor{DARKBLUE}{\sffamily #1}}
\newcommand{\Mag}[1]{\textcolor{DARKMAGENTA}{\sffamily #1}}
\newcommand{\mathblue}[1]{\textcolor{DARKBLUE}{#1}}

\newcommand{\mathmag}[1]{\textcolor{DARKMAGENTA}{#1}}

\newcommand{\YT}[1]{\textcolor{DARKBLUE}{\sffamily [YT: #1]}}

\acrodef{GW}{gravitational wave}
\acrodef{PBH}{primordial black hole}
\acrodef{ABH}{astrophysical black hole}
\acrodef{GWTC-3}{Gravitational Wave Transient Catalog 3}
\acrodef{O3}{third observing run}
\acrodef{QCD}{quantum chromodynamics}

\begin{document}

%\preprint{}
%\preprint{XXX}
%\preprint{YYY}

\reportnum{RUP-22-16}

\title{Effective inspiral spin distribution of primordial black hole binaries}



\author[0000-0002-9579-5787]{Yasutaka Koga}
\email{koga.yasutaka.k2@f.mail.nagoya-u.ac.jp}
%\thanks{}
\affiliation{Division of Particle and Astrophysical Science, Graduate School of Science, Nagoya University, Nagoya 464-8602, Japan}

\author{Tomohiro Harada}
\email{harada@rikkyo.ac.jp}
%\thanks{}
\affiliation{Department of Physics, Rikkyo University, Toshima, Tokyo 171-8501, Japan}

\author[0000-0001-6199-7033]{Yuichiro Tada}
\email{tada.yuichiro.y8@f.mail.nagoya-u.ac.jp}
%\thanks{}
\affiliation{Institute for Advanced Research, Nagoya University, Furocho Chikusaku Nagoya, Aichi 464-8601 Japan}
\affiliation{Division of Particle and Astrophysical Science, Graduate School of Science, Nagoya University, Nagoya 464-8602, Japan}
\affiliation{Theory Center, IPNS, KEK, 
1-1 Oho, Tsukuba, 
Ibaraki 305-0801, Japan}
%\affiliation{Graduate University for Advanced Studies (Sokendai), Tsukuba 305-0801, Japan}

\author{Shuichiro Yokoyama}
\email{shu@kmi.nagoya-u.ac.jp}
%\thanks{}
\affiliation{Kobayashi Maskawa Institute, Nagoya University, Chikusa, Aichi 464-8602, Japan}
\affiliation{Kavli IPMU (WPI), UTIAS, The University of Tokyo, Kashiwa, Chiba 277-8583, Japan}

\author{Chul-Moon Yoo}
\email{yoo@gravity.phys.nagoya-u.ac.jp}
%\thanks{}
\affiliation{Division of Particle and Astrophysical Science, Graduate School of Science, Nagoya University, Nagoya 464-8602, Japan}

%\author{}
%\thanks{}


%\date{\today}

\begin{abstract}
We investigate the probability distribution of the effective inspiral spin, the mass ratio, and the chirp mass of \ac{PBH} binaries, incorporating the effect of the critical phenomena of gravitational collapse.
As a leading order estimation, each binary is assumed to be formed from two \acp{PBH} that are randomly chosen according to the probability distribution of single \acp{PBH}.
We find that, although the critical phenomena can lead to large spins on the low-mass tail, the effective inspiral spin of the binary is statistically very small, $\sqrt{\langle\chi_{\mr{eff}}^2\rangle}=8.41\times10^{-4}$.
We also see that there is almost no anti-correlation between the effective inspiral spin and the mass ratio, which can be inferred from observations. 
\end{abstract}\acresetall

%\pacs{
%04.20.-q, 04.40.Nr, 98.35.Mp
%04.20.-q, 04.70.-s, 04.70.Bw, 95.35.+d, 97.60.Lf, 98.80.-k, 98.80.Cq

%\maketitle

%\section{}
%\subsection{}
%\subsubsection{}

%\tableofcontents

%%%sec
\section{Introduction}
\label{sec:introduction}

%\Mag{On the observational side, PBH scenarios for the black hole binary coalescence events
%have been extensively discussed~\cite{Bird:2016dcv,Clesse:2016vqa,Sasaki:2016jop}.}
The %current 
success in the direct observation %observations 
of gravitational waves from %black hole 
compact binary coalescence %\Blue{s} 
have %\sout{been} 
provided us with much information %of black holes.
about astrophysics and cosmology.
It reveals the abundant existence of massive black holes (see \ac{GWTC-3}~\citep{gwtc3} for the latest data up to the end of LIGO--Virgo's \ac{O3}), shows us the detail dynamics of kilonova~\citep{Metzger:2019zeh} and the properties of high-density nuclear matter~\citep{LIGOScientific:2018cki}, constrains theories of modified gravity \citep{LIGOScientific:2018dkp, LIGOScientific:2021sio}), will be possibly used to measure the Hubble constant of the universe~\citep{LIGOScientific:2019zcs}, etc. However, the origin of the source compact objects (black holes in particular) is not fully clear yet, represented by relatively large masses of black holes compared to the known astrophysical models of the black hole formation (e.g., see the implications of the $150~{\rm M}_\odot$ binary black hole merger~\citep{LIGOScientific:2020ufj}) and the compact binary merger in the low-mass gap~\citep{LIGOScientific:2020zkf}.
Candidate events of subsolar masses, which cannot be stellar black holes, are also reported~\citep{Phukon:2021cus}.
In addition to the total mass of the binary, the effective inspiral spin $\chi_\eff=(a_1\cos\theta_1+qa_2\cos\theta_2)/(1+q)$ and the mass ratio $q=M_2/M_1$ are other important characteristics of binaries to identify their origins, where $M_{i}$, $a_{i}$, and $\theta_{i}$ with $i=1, 2$ are individual masses, individual dimensionless Kerr (spin) parameters, and the angles of individual spins with respect to the orbital angular momentum, respectively.
%where $M_{1/2}$, $a_{1/2}$, and $\theta_{1/2}$ are the mass, the dimensionless Kerr (spin) parameter, and its angle to the inspiral angular momentum of the primary/secondary black hole \har{the binary?} \YT{it is the angle of each BH, so it's not a (characteristics of) the binary}
%\har{Rephrased: where $M_{i}$, $a_{i}$, and $\theta_{i}$ with $i=1, 2$ are individual masses, individual dimensionless Kerr (spin) parameters, and the angles of individual spins with respect to the orbital angular momentum, respctively.}.
The positive (negative) $\chi_\eff$ indicates the alignment (anti-alignment) of their spins.
\citet{Callister_2021} and \citet{theligoscientificcollaboration2021population} found broad distributions of $\chi_\eff$ and $q$ in the observational data (\ac{GWTC-3}), allowing the negative $\chi_\eff$ in the posterior particularly for less hierarchical ones $q\sim1$ (note that there are only two candidate events~(GW191109\_010717 and GW200225\_060421) with significant support, though~\citep{theligoscientificcollaboration2021population}).
Such a spin anti-alignment is counterintuitive because progenitors' spins are expected to be nearly aligned with their orbital angular momentum if they are isolated.
They also remarkably reported
%Black hole binaries can be characterized by their effective \Mag{inspiral} spin, mass ratio, and chirp mass.
%In Ref.~\cite{Callister_2021}, the distribution of the effective \Mag{inspiral} spin and mass ratio was shown by analyzing the observational data of the gravitational waves.
%According to the analysis, the binaries are broadly distributed in the mass ratio, $0.2\lesssim q\leq1$, where $q=M_2/M_1$ is the ratio of the secondary mass $M_2$ to the primary mass $M_1$.
%The effective inspiral spin is also distributed within the width, $|\chi_{\mr{eff}}|\lesssim 0.3$.
%Remarkably, the authors found 
a tendency of anti-correlation between the mean value of $\chi_{\mr{eff}}$ and the mass ratio $q$, that is, the average $\chi_{\mr{eff}}$ has a slightly larger positive value for smaller $q$, a hierarchical mass configuration.
They noted that this tendency is also in an opposite sense to the standard astrophysical models (see \cite{Callister_2021} and references therein).
%The analysis of Ref.~\cite{theligoscientificcollaboration2021population} also shows a broad distribution in $\chi_{\mr{eff}}$ with a slight tendency for positive $\chi_{\mr{eff}}$.
%\Mag{Thus, the information on the spin as well as the mass of the black holes can provides important insights into the black hole formation scenario.}

%Primordial black holes (PBHs) 
In addition to the \ac{ABH}, the so-called \ac{PBH} has been also extensively discussed as a candidate of merger black holes (see \citet{Bird:2016dcv}, \citet{Clesse:2016vqa}, and \citet{Sasaki:2016jop} for the first proposals).
\Acp{PBH} are hypothetical black holes formed in the early universe without introducing massive stars contrary to the ordinary \acp{ABH}~\citep{Zeldovich:1967lct, Hawking:1971ei,Carr:1974nx}.
%%%%%%%% by SY %%%%%%%%%%%%%
%, such as the radiation-dominated era
%As they are formed through non-astrophysical processes, the study of them provides unique probes of %various physics.
%For example, the possibility that PBHs occupy a fraction of the dark matter density in the universe has %been extensively investigated so far (See Ref.~\cite{Carr_2021} and references therein).
%%%%%%%%%%%%%%%%%%%%%%%%%%%%
While many formation mechanisms have been proposed, one main scenario
%One of the formation mechanisms 
is the collapse of an overdense region of the universe. 
%In the early inflationary phase, quantum fluctuations are produced at sub-horizon scales and, after horizon exit, considered to become classical density perturbations at super-horizon scales.
%In the subsequent radiation dominated phase, the cosmological perturbations enter the horizon and start gravitational contraction.
%Then, the perturbations with an amplitude greater than a threshold will collapse into PBHs.
If primordial density perturbations $\delta$ generated by cosmic inflation are large enough and exceed a threshold value $\delta_\uth$, they can gravitationally collapse directly into black holes soon after their horizon reentry \citep{Carr:1975qj, Nadezhin:1978, Harada:2013epa}.
Since the \ac{PBH} masses are roughly given by the Hubble masses at their formation times, they can be distributed in the very broad range, $10^{-5}$--$10^{50}\,\si{g}$, including both massive ones and subsolar ones (\acp{PBH} with masses smaller than $10^{15}\,\si{g}$ are considered to have evaporated away by the present epoch due to the Hawking radiation)~\citep{carr}.

If we focus on PBHs formed in the radiation dominated era, the spins of \acp{PBH} have been thought to be small typically because they originate from the almost spherically symmetric contraction of the Hubble patch.
Recently, the spin distribution of PBHs has been extensively studied~\citep{Chiba:2017rvs, Harada:2017fjm, Mirbabayi:2019uph, He:2019cdb, Flores:2021tmc, Chongchitnan:2021ehn, Eroshenko:2021sez}.
In particular, \citet{Luca_2020} and subsequently %, by %Harada et al.~
\citet{Harada_2021} carefully investigated it
based on the so-called peak theory~\citep{Bardeen:1985tr} of the cosmological perturbation.
\citet{Harada_2021} found that the root mean square of the initial value of the nondimensional Kerr parameter is given by a form proportional to $(M/M_H)^{-1/3}$ with a typically small numerical factor of $\mathcal{O}(10^{-3})$, where $M$ and $M_H$ are the PBH mass and the Hubble mass at the formation, respectively.
This result implies that, for ordinary formation of PBHs such that $M\sim M_H$, the spin parameter is very small as $\sim10^{-3}$ in fact.
However, according to numerical simulations, the so-called critical phenomena have been reported for the \ac{PBH} formation on the other hand~\citep{Evans_1994,Niemeyer:1997mt,Niemeyer:1999ak,Yokoyama:1998xd,Green:1999xm,Koike:1995jm,Musco:2004ak,Musco:2008hv,Musco:2012au,Escriva_2020}. %\YT{enough ref?}\CY{I added related papers considering critical phenomena of radiation fluid and PBH. maybe too many...}
That is, the resultant \ac{PBH} mass is not necessarily given by the Hubble mass but in a scaling relation $M\propto M_H(\delta-\delta_\uth)^{\kappa}$ with the universal power $\kappa\simeq0.36$. Therefore, the mass can be arbitrarily small as $M\ll M_H$ for $\delta\sim\delta_\uth$ and rapidly spinning PBHs could be formed in that case.
%The spins of black holes can be measured through, for example, observation of gravitational waves from %black hole binaries in terms of the effective spin, a weighted sum of the components of the black hole %spins parallel to the orbital angular momentum.
%%%%%%%%%%% moved to here by SY %%%%%%%%%%
Furthermore, \acp{PBH} basically have no correlation with each other because they are separated farther than the Hubble scale at their formation time, which makes the spin anti-alignment more natural for the \ac{PBH} binaries.
%%%%%%%%%%%

In this paper, based on the above-mentioned observational and theoretical backgrounds, we investigate probability distribution of the effective inspiral spin $\chi_{\mr{eff}}$, mass ratio $q$, and chirp mass $\calM$ of PBH binaries, taking account of the critical phenomena.
In particular, as the mass-spin anti-correlation $\sqrt{\braket{a_*^2}}\propto(M/M_H)^{-1/3}$ is reported, it is interesting to see the correlation between $\chi_\eff$ and $q$ as found in observations.
%As the first step of this kind of works, 
While several scenarios have been proposed for binary formation of \acp{PBH} (see, e.g., \citet{Nakamura:1997sm} and \citet{Bird:2016dcv}),
we %consider a simplest model such 
simply assume that each binary is formed from two randomly chosen PBHs, %according to the probability distribution of single PBHs.
considering it as a leading order approximation in any case.
%Especially, we take the effect of the critical phenomena of gravitational collapse into account.
%Since, as suggested in Ref.~\cite{Harada_2021}, the critical phenomena can produce rapidly spinning PBHs, it is interesting to investigate the effective spin distribution of binaries incorporating the effect of the critical phenomena.
Therefore, the probability distribution of the single \ac{PBH}~\citep{Harada_2021} can be straightforwardly extended to the binary system.

This paper is organized as follows.
In Sec.~\ref{sec:distribution-pbh}, we derive the probability distribution of single PBHs incorporating the effect of the critical phenomena of gravitational collapse.
For simplicity, we assume an almost monochromatic power spectrum of the density fluctuation that will collapse into a PBH.
In Sec.~\ref{sec:distribution-binary}, we derive and numerically estimate the probability distribution of PBH binaries formed from two randomly chosen PBHs.
The conclusion is given in Sec.~\ref{sec:conclusion}.
We use units in which $c=1$.

%%%sec
\section{Distribution of single PBHs}
\label{sec:distribution-pbh}

Let us discuss the statistics of single PBHs in this section. We focus on the PBH formation via the collapse of overdensities in the radiation-dominated universe. Due to the charge-neutrality of the universe, PBHs are basically assumed neutral electromagnetically, and hence the no-hair theorem tells us that PBHs are characterized only by their masses $M$ and spin vectors $a^i=a(\sin\theta\cos\phi,\sin\theta\sin\phi,\cos\theta)^T$, where we employ the dimensionless definition $a^i=S^i/GM^2$ with the angular momentum $S^i$ and its norm $a$ is often called \emph{Kerr parameter}.
The PBH statistics is accordingly dictated by the probability distribution of their characteristics,
\bae{\label{eq:single-PBH-P}
    P(a,M,\theta,\phi)\dd{a}\dd{M}\dd{\theta}\dd{\phi},
}
and the statistical-isotropy assumption restricts its form as
\bae{
\label{eq:isotropic-dist}
    P(a,M,\theta,\phi)\dd{a}\dd{M}\dd{\theta}\dd{\phi}=\frac{1}{4\pi}P(a,M)\dd{a}\dd{M}
    \dd{\mu}
    \dd{\phi},
}
with $\mu = \cos \theta$, which consistently gives the distribution of the Kerr parameter and mass as
\bae{\label{eq: def of PaM}
    P(a,M)=\int P(a,M,\theta,\phi)\dd{\theta}\dd{\phi}.
}
We below derive this distribution $P(a,M)$.

%%%%%%%%%%%%%%%%%%%%%%%%%%%%%%%%%
\begin{comment}
\YT{In order to define relevant quantities as well, a quick overview about the PBH analysis can be shown first}
A single PBH has spin
$a^i=a(\sin\theta\cos\phi,\sin\theta\sin\phi,\cos\theta)$ and
mass $M$\red{, where $a$ is its Kerr parameter}. %\YT{define $a$}
Their probability distribution can be expressed as
\begin{equation}
    P(a,M,\theta,\phi)\dd{a}\dd{M}\dd{\theta}\dd{\phi}.
\end{equation}
Assuming isotropic distribution of the spin direction, %we have
\Blue{it should have the form of}
\begin{equation}
\label{eq:angle-separation}
    P(a,M,\theta,\phi)\dd{a}\dd{M}\dd{\theta}\dd{\phi}
    %=P(a,M)P(\theta,\phi)dadMd\theta d\phi
    =\frac{1}{4\pi}P(a,M)\dd{a}\dd{M}\sin\theta\dd{\theta}\dd{\phi},
\end{equation}
\Blue{where the distribution of the spin magnitude and mass $P(a,M)$ is defined by
\bae{
    P(a,M)=\int P(a,M,\theta,\phi)\dd{\theta}\dd{\phi}.
}}
We derive the distribution function of the spin and the mass, $P(a,M)$, in the following.
\end{comment}
%%%%%%%%%%%%%%%%%%%%%%%%%%%%%%

%%%subsec
\subsection{Spin distribution}

PBHs are supposed to be formed by the collapse of rare highly-overdense regions. According to the \emph{peak theory}~\citep{Bardeen:1985tr},
if the density contrast $\delta$ follows the Gaussian distribution and characterized by
an almost monochromatic power spectrum $\calP_\delta(k)\approx\sigma_0^2k_0\delta(k-k_0)$ with some scale $k_0$ as we assume throughout this paper, the spatial profile of such ``high peaks" of the Gaussian random field is known to be typically spherically symmetric and given by~\citep{Yoo_2018}
\bae{\label{eq: peak profile}
    \delta_\pk(r)\simeq\nu\sigma_0\frac{\sin k_0r}{k_0r},
}
with a (normalized) random Gaussian parameter $\nu$ following the distribution $P_\uG(\nu)=\frac{1}{\sqrt{2\pi}}\ee^{-\nu^2/2}$. 
The peak extremum is put at the origin $r=0$ without loss of generality.
The PBH characteristics (mass and spin as well as whether they are really formed or not) are parametrized by this $\nu$ parameter.
In this subsection, we first review the spin distribution determined by $\nu$, following \citet{Harada_2021} (see also \citet{Heavens_1988} and \citet{Luca_2019}).
Note that, although peaks of a Gaussian random field do not necessarily obey a Gaussian distribution, we assume a Gaussian distribution of $\nu$ as an approximation.
The validity is discussed in Appendix~\ref{sec:peak-number}, and the appropriate normalization factor for the PBH case is given in the next subsection.

Though the typical peak profile is almost spherically symmetric, a slight deviation from an exactly monochromatic spectrum can cause a tidal torque introducing a spin to a PBH.
In the peak theoretical approach, \citet{Harada_2021} revealed that the normalized spin parameter $h$, which is defined in Eq.~\eqref{eq:def-h} of Appendix~\ref{sec:def-h}, is related to $a$ and $\nu$ as
\beae{\label{eq: def of h}
    &h=\frac{a}{C(M,\nu)}, \\
    &C(M,\nu)=
    3.25\times 10^{-2}\sqrt{1-\gamma^2}\sigma_0\left(\frac{M}{M_H}\right)^{-1/3}\!\left(\frac{\nu}{10}\right)^{-1}\!,
}
and follows the universal distribution\footnote{
In the recent work~\citep{Luca_2019}, %by De Luca et al., %the 
another fitting distribution function is given %by
as
\bme{
    P_h(h)\dd{h} \\ 
    =\exp[-2.37-4.12\ln h-1.53(\ln h)^2-0.13(\ln h)^3 ]\dd{h}.
}
%in Eq.~(6.30) as a fitting expression.
However, since it is singular for $h\to0$, we here adopt the %other 
original and regular fitting expression~\eqref{eq:Ph-Heavens} given by %Heavens and Peacock~
\citet{Heavens_1988}.
}
\bae{\label{eq:Ph-Heavens}
    &P_h(h)\dd{h}=563h^2 \nonumber \\ 
    &\times\exp\bqty{-12h+2.5h^{1.5}+8-3.2(1500+h^{16})^{1/8}}\dd{h},
}
which is a fitting formula found %in Ref.~
by \citet{Heavens_1988} (note that it is normalized so that $\int_0^\infty P_h(h)\dd{h}=1$).
Here $M$ is the total mass of the collapsing fraction, $M_H$ is the horizon mass at the horizon reentry of the overdense region, and $\gamma\coloneqq\sigma_1^2/(\sigma_0\sigma_2)$ with $\sigma_j^2\coloneqq\int\dd{\ln k} k^{2j}\calP_\delta(k)$ characterizes the width of the power spectrum of the density contrast ($\gamma=1$ for an exactly monochromatic spectrum). 
%\SY{What value is used in this paper for $\gamma$?}.
Throughout this paper, we assume $\gamma=0.85$.
Given $M$ and $\nu$, the PBH spin distribution 
can be deduced basically from this formula.
See Appendix~\ref{sec:def-h} for the brief introduction of $h$ and $C(M,\nu)$.

One should note that a PBH is not necessarily formed for a given $\nu$ and thus the PBH formation condition should be imposed to obtain the spin distribution of PBHs.
For an almost monochromatic spectrum (and thus for an almost uniform typical peak profile~\eqref{eq: peak profile}), it is justified to judge the PBH formation just by whether $\nu$ exceeds some threshold value $\nu_\uth$ (see, e.g., \citet{Germani:2018jgr}).
We basically neglect the spin dependence of $\nu_\uth$
%\SY{Do we need to mention, e.g., Ref.~\cite{Baumgarte:2016xjw}?}\YK{Added the footnote.}
but just take account of the fact that $a>1$ is not allowed for a BH.\footnote{From the investigations of the critical phenomena in asymptotically flat cases~\citep{Baumgarte:2016xjw,Gundlach:2016,Celestino:2018}, it is implied that the spin
dependence of the threshold is weak if the initial matter distribution is nearly spherically symmetric. Since the typical profile of the density perturbation which collapses into a PBH is almost spherically symmetric, the assumption~\eqref{eq: spin dependence of nuth} would not greatly affect the resulting distribution of single PBHs.} 
That is, we adopt the following simplified spin dependence in this paper:
\bae{\label{eq: spin dependence of nuth}
    \nu_\uth=\bce{
        \mathrm{const.} & \text{for $0\leq a\leq1$}, \\
        \infty & \text{otherwise}.
    }
}
Therefore, the distribution of the PBH Kerr parameter $a$ given $M$ and $\nu$ is simply obtained by the change of the variable $h\to a=Ch$ for $0\leq h\leq 1/C$ as
\bae{
    P(a\mid M,\nu)\dd{a}=\frac{P_h\bigl(a/C(M,\nu)\bigr)}{C(M,\nu)N_a(M,\nu)}\dd{a},
}
with the normalization factor
\bae{
    N_a(M,\nu)\coloneqq\int_0^{1/C(M,\nu)}P_h(h)\dd{h},
}
to ensure $\int_0^1P(a\mid M,\nu)\dd{a}=1$.
The simplified spin dependence~\eqref{eq: spin dependence of nuth} does not make a problem practically because the typical PBH spin is quite small as $a\ll1$ as we will see below.

\subsection{Critical behavior}
\label{sec:critical-behavior}

The PBH mass $M$ also depends on $\nu$. It is roughly equivalent to the horizon mass $M_H$ at the horizon reentry of the overdense region, but in detail, it is often assumed to follow the scaling relation, %\YT{the character $\beta$ is confusing with the PBH abundance?}\YK{the power is changed as $\beta\to\bar{\beta}$}
\bae{\label{eq:critical-behavior}
    M(\nu)=KM_H(\nu\sigma_0-\nu_\uth\sigma_0)^{\kappa},
}
through which the PBH mass can be understood as a function of $\nu$. Here $\kappa \simeq 0.36$ is a universal power and $K$ is a weakly profile-dependent coefficient~\citep{Evans_1994,Escriva_2020}.
Since $K$ is of order unity in any case, we here set $K=1$ for simplicity.
Once $M$ is related to $\nu$, the joint probability $P(a,M)$ for a PBH can be calculated as
\bae{\label{eq:PaM}
    P(a,M)\dd{a}\dd{M}=P(a\mid M(\nu),\nu)P_\nu(\nu)\dd{a}\dd{\nu},
}
where
\bae{\label{eq: Pnu for PBH}
    P_\nu(\nu)=\sqrt{\frac{2}{\pi}}\frac{\ee^{-\nu^2/2}}{\erfc(\nu_\uth/\sqrt{2})},
}
is the Gaussian distribution of $\nu$ \emph{for a PBH}, i.e., it is defined only for $\nu>\nu_\uth$ and normalized as $\int_{\nu_\uth}^\infty P_\nu(\nu)\dd{\nu}=1$ with $\erfc$ denoting the complementary error function.
For the derivation of Eq.~\eqref{eq: Pnu for PBH}, see Appendix~\ref{sec:peak-number}.

\begin{figure}
    \centering
    \includegraphics[width=0.95\hsize]{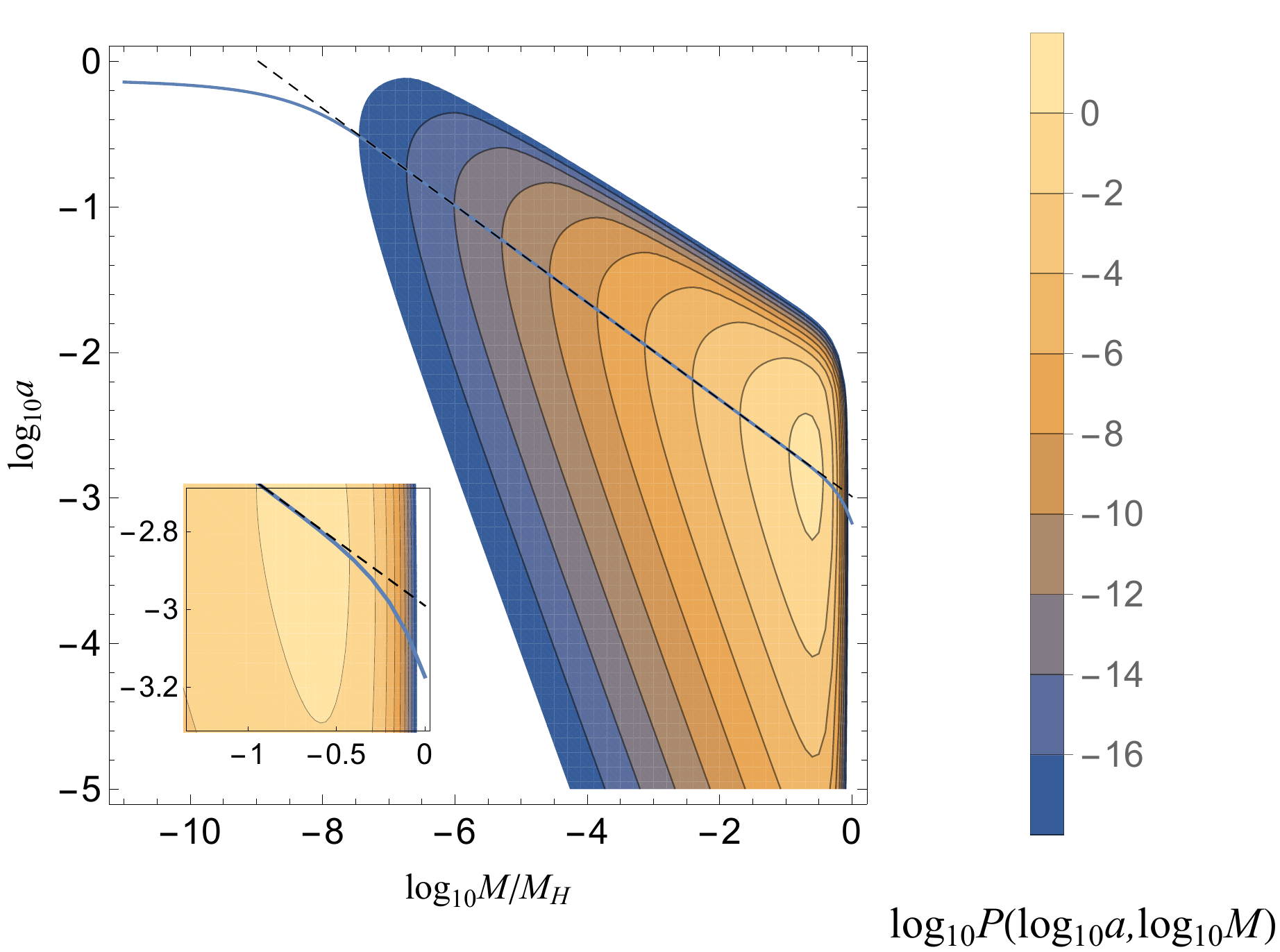}
    \caption{
    A contour plot
    of $\log_{10}P(\log_{10}a,\log_{10}M)$ for $\nu_\uth=10$ and $\sigma=0.192$ %which correspond to $f_\PBH\sim0.1\%$ for $M_H\sim M_\odot$.
    (see discussions in the next subsection).
    The solid line shows the expected value $\langle a\rangle$ for each $M$, while the dashed line is its power-law fitting $\propto (M/M_H)^{-1/3}$.}
    \label{fig:PaM-sgm0192}
\end{figure}

The plot of 
$P(\log_{10}a,\log_{10}M)=P(a,M)(\ln {10})^2aM$
%$P(\log_{10}a,\log_{10}M)=P(a,M)(\log_{10})^2aM$
is shown in Fig.~\ref{fig:PaM-sgm0192} for $\nu_\uth=10$ and $\sigma=0.192$, which correspond to %$f_\PBH\sim0.1\%$ for $M_H\sim M_\odot$.
$0.1\%$ fraction of dark matter with $M_H\sim M_\odot$ as will be discussed in the next subsection.
The \ac{PBH} spin and mass are mostly distributed in the range of $10^{-4}\lesssim a\lesssim 10^{-3}$ %$0.0001\lesssim a\lesssim 0.004$
and $0.1\lesssim M/%M_{k_0}
M_H\lesssim0.4$. % with about $90\%$ probability.
The expected value of $a$ for each $M$ defined by $\langle a(M)\rangle=\int_0^1 aP(a,M)\dd{a}/\int_0^1 P(a,M)\dd{a}$ is also plotted.
The power law, $\langle a\rangle\propto M^{-1/3}$, can be seen in the range, $10^{-8}\lesssim M/M_{%k_0
H}\lesssim 0.3$, %\YT{what's the reason for the upper bound?}\YK{comment added in the end of this paragraph}, 
as expected from the normalization $C(M,\nu)$~\eqref{eq: def of h} (see \citet{Harada_2021}).
%Roughly speaking, 
This anti-correlation between $a$ and $M$ is because,
%a more massive \ac{PBH} corresponds to a higher peak which tends to be more spherically symmetric.
as we can see from Eqs.~\eqref{eq:AppA-Sref},~\eqref{eq:AppA-Aref}, and~\eqref{eq:AppA-Mta-f}, the magnitude of the total angular momentum of the collapsing fraction scales as $S_\mr{ref}\propto(M/M_H)^{5/3}$ and the corresponding Kerr parameter scales as $a\propto A_\mr{ref}=S_\mr{ref}/(GM^2)\propto (M/M_H)^{-1/3}$. %\YT{is it correct?}\YK{partially? even if a peak is almost spherically symmetric, the resulting PBH can have nonzero $a$ if $M/M_H<<1$. Even if the total angular momentum of the mass inside Habble horizon is zero, the fraction $M/M_H$ collapsing into a PBH can have nonzero angular momentum.}
Also note that %the 
this power law is violated for the much smaller mass, $M/M_{%k_0
H}\lesssim10^{-8}$, corresponding to the limit, $\langle a\rangle\to1$, because of our assumption that a peak of the density fluctuation with $a>1$ will not collapses into a PBH.
The violation for $M/M_H\gtrsim0.3$ appears because the factor $\nu(M)^{-1}$ in $C(M,\nu(M))$ is not constant in this range, while it is almost constant, $\nu(M)^{-1}\simeq\nu_\uth^{-1}$, for $M/M_H\lesssim0.3$.

\subsection{PBH abundance}
\label{sec:PBH-abundance}

In order to concretely specify the parameters, let us also review the current PBH abundance.
The normalization of $\nu$'s distribution for a PBH~\eqref{eq: Pnu for PBH} implies that a PBH can be formed with the probability $\int_{\nu_\uth}^\infty P_\uG(\nu)\dd{\nu}=\erfc(\nu_\uth/\sqrt{2})/2$ at each Hubble patch.
Uniformly approximating the PBH mass by the horizon mass for simplicity, the ratio $\beta$ of the PBH energy density to that of the background radiation at their formation time is hence given by that probability:
\bae{
    \beta=\frac{1}{2}\erfc\qty(\frac{\nu_\uth}{\sqrt{2}}).
}
After their formation, PBHs behave as non-relativistic matters and their energy density decays as $\propto \mathfrak{a}^{-3}$ where $\mathfrak{a}$ is the scale factor of the universe.
Accordingly, one can calculate the ratio of the current PBH energy density $\rho_{\PBH,0}$ to that of the total cold dark matters $\rho_{\DM,0}$ and it reads (see, e.g., \citet{Tada:2019amh})
\bae{\label{eq: fPBH}
    &f_\PBH=\frac{\rho_{\PBH,0}}{\rho_{\DM,0}} \nonumber \\
    &\sim\pqty{\frac{\beta}{1.8\times10^{-9}}}\pqty{\frac{\Omega_\DM h^2}{0.12}}^{\!-1}\pqty{\frac{g_*}{10.75}}^{\!-1/4}\pqty{\frac{M}{M_\odot}}^{\!-1/2}\!,
}
where $\Omega_\DM h^2\simeq0.12$ is the current density parameter of the total cold dark matters~\citep{Planck:2018vyg}, $g_*\simeq10.75$ is the effective degrees of freedom for energy density of the radiation fluid at the formation time of solar-mass PBHs, and $M_\odot\simeq\SI{2e33}{g}$ is the solar mass.
Here we approximated the effective degrees of freedom for entropy density by those for energy density, $g_{*s}\approx g_*$, throughout the history and assumed that PBHs were formed at around the time when $k_0$ reentered the horizon.
It has been implied that $f_\PBH\sim0.1\%$ to account for the merger rate of BH binaries inferred LIGO (see, e.g., %Sasaki:2018dmp} 
%\YT{other refs?} \Mag{$\to$ Refs.~
\citet{Ali-Haimoud:2017rtz} and \citet{Vaskonen:2019jpv}).
From the formula~\eqref{eq: fPBH}, one sees that this abundance corresponds to $\nu_\uth\sim10$ for $M\sim M_\odot$.
Below, we will employ this value of $\nu_\uth$.

One can also infer the perturbation amplitude $\sigma_0$ from the value of $\nu_\uth$.
The PBH formation is often judged by using the so-called compaction function $\calC(r)$ which is defined by Eq.~(4.28) in \citet{Shibata:1999zs} or Eq.~(6.33) in \citet{Harada:2015yda} for 
the constant-mean-curvature slicing. 
If the maximum $\calC_\um\coloneqq\max\Bqty{\calC(r)\mid r}$ exceeds the threshold $\calC_{\um,\uth}\sim2/5$, which has been suggested by fully non-linear numerical simulations~\citep{Shibata:1999zs,Harada:2015yda,Musco:2018rwt,Germani:2018jgr}, for some overdense region, that region is supposed to form a PBH.
Assuming the peak profile~\eqref{eq: peak profile}, the maximum $\calC_\um$ corresponds to the central value $\delta_\pk(r=0)$ by $\calC_\um\simeq(5/24)\delta_\pk(0)$ (see \citet{Harada_2021} for details).
In order for $\nu_\uth$ to correspond to $\calC_\uth$, the perturbation amplitude $\sigma_0$ should be given by $\sigma_0\simeq(24/5)(\calC_{\um,\uth}/\nu_\uth)\simeq0.192$.

\section{Distribution of PBH binaries}
\label{sec:distribution-binary}

As only one \ac{PBH} forms in one Hubble patch, basically \acp{PBH} have no correlation with each other before their formations. They randomly form in space and some of them make binaries by gravitationally catching each other through several proposed scenarios such as free falls of two near \acp{PBH} in the early universe~\citep{Nakamura:1997sm} or gravitational captures in galactic halos in the late universe~(see, e.g., \citet{Bird:2016dcv}). Anyway two \acp{PBH} forming a binary can be assumed to be chosen randomly.

A black hole binary system %can be 
seen by its merger \acp{GW} is characterized by the chirp mass $\mathcal{M}$, the mass ratio $q$, and the effective inspiral spin $\chi_{\mr{eff}}$ defined by
\beae{\label{eq:chi-q}
    \calM&=\frac{(M_1M_2)^{3/5}}{(M_1+M_2)^{1/5}}\in(0,\infty), \\
    q&=\frac{M_2}{M_1}\in(0,1], \\
    \chi_{\mr{eff}}&=\frac{a_1\mu_1+qa_2\mu_2}{1+q}\in[-1,1],
}
respectively.
Here, the quantities with the subscripts $1$ and $2$ are those of the primary and secondary PBHs, respectively.
The polar angles of spins, $\theta_1$ and $\theta_2$, are taken so that the axis coincides with the orbital angular momentum, ${\bf L}$.
As mentioned, the primary (PBH1) and secondary (PBH2) \acp{PBH} are assumed to be chosen randomly according to their probability distribution, Eqs.~\eqref{eq:single-PBH-P} and \eqref{eq:PaM}.
Moreover, for simplicity, we assume that the mass and spin angular momenta of PBHs are constant during the formation process of two isolated PBHs to a binary.
Thus we can straightforwardly derive the probability distribution of \ac{PBH} binaries, $P(\calM,q,\chi_{\mr{eff}})\dd{\calM}\dd{q}\dd{\chi_{\mr{eff}}}$ from this single-\ac{PBH} distribution.
%\YT{how the circular-orbit assumption affect the result? Does it change the first assumption that a \ac{PBH} binary is characterized only by $\calM$, $q$, and $\chi$?}

%In the following, we derive the probability distribution of PBH binaries, $P(\calM,q,\chi)\dd{\calM}\dd{q}\dd{\chi}$, including the effect of the critical phenomena.
%We assume that the primary (PBH1) and secondary (PBH2) PBHs are randomly chosen according to their probability distribution, Eqs~\eqref{eq:single-PBH-P} and~\eqref{eq:PaM}, and are in circular orbits.
%Then, the configuration of the PBH binary is determined by its intrinsic parameters, $(a_i,M_i,\Mag{\mu_i},\phi_i)$ $(i=1,2)$, i.e., the parameters of PBH1 and PBH2.

%%%subsec
%\subsection{Intrinsic Parameter Distribution}
\medskip

%Since the PBH1 and PBH2 are randomly chosen according their probability distribution, the probability distribution of the intrinsic parameters of the binary, ${\bf x}=(a_1,a_2,M_1,M_2,\Mag{\mu_1},\Mag{\mu_2},\phi_1,\phi_2)$, is given by
Thanks to the independence of PBH1 and PBH2, one first obtains the joint probability distribution of their intrinsic parameters $\bfw=(a_1,a_2,M_1,M_2,\mu_1,\mu_2,\phi_1,\phi_2)$ as a direct product of each probability,
\bae{
    P(\bfw)\dd{\bfw}
    = \frac{2}{(4 \pi)^2}\prod_{i=1}^2
    P(a_i,M_i)\dd{a_i}\dd{M_i}\dd{\mu_i}\dd{\phi_i},
}
where we have normalized the PDF, $P(\bfw)$, so that its integration over $a_i\in[0,1]$, $0<M_2\le M_1<\infty$, %$\theta_i\in(0,\pi)$, 
$\mu_i \in[-1,1]$, and $\phi_i\in[0,2\pi)$ becomes unity.
Note that the isotropy assumption~\eqref{eq:isotropic-dist} has been also used.
According to the argument on the critical behavior in Sec.~\ref{sec:critical-behavior}, the distribution of the variables $a_i$ and $M_i$ is read as that of $a_i$ and $\nu_i$ by using Eq.~\eqref{eq:PaM}.
Thus, we also have
\bme{
    P(\bfw^\prime)\dd{\bfw^\prime} \\
    =\frac{1}{8\pi^2}\prod_{i=1}^2
    P(a_i\mid M_i(\nu_i),\nu_i)P_\nu(\nu_i)\dd{a_i}\dd{\nu_i}\dd{\mu_i}\dd{\phi_i},
}
for the variables, ${\bfw^\prime}=(a_1,a_2,\nu_1,\nu_2,\mu_1,\mu_2,\phi_1,\phi_2)$.
Here, $\nu_i$ is the peak value of each density fluctuation that forms each PBH of the binary.
%%%%% moved by SY 0601 %%%%%%%%%%%%%%
Noting the critical behavior~\eqref{eq:critical-behavior}, the Jacobian reads
%\begin{eqnarray}
\bae{
    J_{
    ww^\prime}&=\abs{\dv{\bfw}{\bfw^\prime}} \nonumber \\
    &=  %K^2 
    M_{H}^2\kappa^2\sigma_0^2(\nu_1\sigma_0-\nu_\uth\sigma_0)^{\kappa-1}(\nu_2\sigma_0-\nu_\uth\sigma_0)^{\kappa-1}\nonumber
    \\
    &=M_1M_2\kappa^2\sigma_0^2\pqty{\frac{M_1}{M_{H}}}^{-1/\kappa}\pqty{\frac{M_2}{M_{H}}}^{-1/\kappa},
%\end{eqnarray}
}
and the probability in $\bfw$ is given by
\bae{
    P(\bfw)&=J_{
    ww^\prime}^{-1}P(\bfw^\prime) \nonumber \\
    &=\frac{1}{8\pi^2J_{
    ww^\prime}}\prod_{i=1}^2P(a_i\mid M_i,\nu(M_i))P_{\nu}(\nu(M_i)),
}
where
\begin{equation}
    \nu(M)=\frac{1}{\sigma_0}\pqty{\frac{M}{M_{H}}}^{1/\kappa}+\nu_\uth.
\end{equation}
%%%%%%%%%%%%%%%%%%%%%%%%%%%%%%%%%

%%%subsec
%\subsection{Distribution of $\calM$, $q$, and $\chi$}

%Let us consider the parameter set $\bfy=(\ln\calM,q,\chi,a_1,a_2,\Mag{\mu_1},\phi_1,\phi_2)$ (we adopt $\ln\calM$ in order for a dimensionless probability) and find the probability distribution $P(\ln\calM,q,\chi)$.
It can be further translated to the parameter set $\bfz=(\calM,q,\chi_{\mr{eff}},a_1,a_2,\mu_1,\phi_1,\phi_2)$.
%We go step by step, first considering the pull back from $\bfw^\prime=(a_1,a_2,\nu_1,\nu_2,\cos\theta_1,\cos\theta_2,\phi_1,\phi_2)$ to $\bfw=(a_1,a_2,M_1,M_2,\cos\theta_1,\cos\theta_2,\phi_1,\phi_2)$.
%Noting the critical behavior~\eqref{eq:critical-behavior}, the Jacobian reads
%\begin{eqnarray}
%   J_{x x^\prime}=\abs{\dv{\bfw}{\bfw^\prime}} 
%   &=&K^2M_{k_0}^2\bar{\beta}^2\sigma_0^2(\nu_1\sigma_0-\nu_\uth\sigma_0)^{\bar{\beta}-1}(\nu_2\sigma_0-\nu_\uth\sigma_%% 0)^{\bar{\beta}-1}\nonumber
%    \\
%    &=&M_1M_2\bar{\beta}^2\sigma_0^2\pqty{\frac{M_1}{KM_{k_0}}}^{-1/\bar{\beta}}\pqty{\frac{M_2}{KM_{k_0}}}^{-1/\bar{\beta}},
%\end{eqnarray}
%and the probability in $\bfx$ is given by
%\begin{equation}
%    P(\bfx)=J_{x x^\prime}^{-1}P(\bfx^\prime)=\frac{1}{8\pi^2J_{x x^\prime}}\prod_{i=1}^2P(a_i\mid %M_i,\nu(M_i))P(\nu(M_i)),
%\end{equation}
%where
%\begin{equation}
%    \nu(M)=\frac{1}{\sigma_0}\pqty{\frac{M}{KM_{k_0}}}^{1/\bar{\beta}}+\nu_\uth.
%\end{equation}
Recalling the definitions of the effective inspiral spin $\chi_{\mr{eff}}$,  the mass ratio $q$, and the chirp mass $\calM$~\eqref{eq:chi-q}, the Jacobian from $\bfw$ to $\bfz$ can be then computed as
\bae{
    J_{
    zw}&=\abs{\dv{\bfz}{\bfw }}=\frac{a_2M_2}{M_1^2(M_1+M_2)}\frac{(M_1M_2)^{3/5}}{(M_1+M_2)^{1/5}} \nonumber \\
    &=\frac{a_2q^{11/5}}{(1+q)^{7/5} %\calM^2
    \calM},
}
where we used the inverse relation
\beae{
    M_1(\calM,q)&=q^{-3/5}(1+q)^{1/5}\calM, \\
    M_2(\calM,q)&=q^{2/5}(1+q)^{1/5}\calM.
}
The probability in $\bfz$ is hence found as
\bae{
    P(
    \bfz)&=J_{
    zw}^{-1}P(\bfw) \nonumber \\
    &=\frac{1+q}{a_2q^2\kappa^2\sigma_0^2\calM}\pqty{\frac{(1+q)^{2/5}\calM^2}{q^{1/5}M_{H}^2}}^{1/\kappa} \nonumber \\
    &\quad\begin{multlined}
        \times\frac{1}{8\pi^2}\prod_{i=1}^2P\Bigl(a_i\relBigm{|} M_i(\calM,q),\nu\qty(M_i(\calM,q))\Bigr) \\
        \times P_{\nu}\Bigl(\nu\qty(M_i(\calM,q))\Bigr).
    \end{multlined}
}

The probability only of $\calM$, $q$, and $\chi_{\mr{eff}}$ is obtained as the integration over the rest variables $a_1$, $a_2$, $\mu_1$, $\phi_1$, and $\phi_2$:
\bae{
    P(\calM,q,\chi_{\mr{eff}})=\int P(
    \bfz)\dd{a_1}\dd{a_2}\dd{\mu_1}\dd{\phi_1}\dd{\phi_2}.
}
Note that the range of $\mu_1$, originally in $(-1,1)$, is now restricted by the other variables, $\chi_{\mr{eff}}$, $q$, $a_1$, and $a_2$, as
\bae{
	\mu_1&=\frac{(1+q)\chi_{\mr{eff}}-qa_2\mu_2}{a_1} \nonumber \\
	&\in\left(\frac{(1+q)\chi_{\mr{eff}}-qa_2}{a_1},\frac{(1+q)\chi_{\mr{eff}}+qa_2}{a_1}\right),
}
because of the range of $\mu_2\in(-1,1)$.
As a result, we have
\bme{
    \mu_1\in\left(\max\left[-1,\frac{(1+q)\chi_{\mr{eff}}-qa_2}{a_1}\right], \right. \\
    \left.\min\left[1,\frac{(1+q)\chi_{\mr{eff}}+qa_2}{a_1}\right]\right).
}
Finally we obtain the %probability distribution function,
expression
\bae{
\label{eq:P-lncaM-q-chi}
    &P(%\ln\calM
    \calM,q,\chi_{\mr{eff}})=
    \frac{1+q}{2q^2\kappa^2\sigma_0^2\calM}\pqty{\frac{(1+q)^{2/5}\calM^2}{q^{1/5}M_{H}^2}}^{1/\kappa} \nonumber \\
    &\begin{multlined}
        \times\int_0^1\dd{a_1}\int_0^1\dd{a_2}\Theta\pqty{T(a_1,a_2,\chi_{\mr{eff}},q)}T(a_1,a_2,\chi_{\mr{eff}},q) \\
        \times\frac{1}{a_1a_2}\prod_{i=1}^2P\Bigl(a_i\relBigm{|} M_i(\calM,q),\nu\qty(M_i(\calM,q))\Bigr) \\
        \times P_{\nu}\Bigl(\nu\qty(M_i(\calM,q))\Bigr),
    \end{multlined}
}
where
\bme{
    T(a_1,a_2,\chi_{\mr{eff}},q)=\min[a_1,qa_2+(1+q)\chi_{\mr{eff}}] \\ 
    +\min[a_1,qa_2-(1+q)\chi_{\mr{eff}}].
}

\begin{figure*}
    \centering
    \includegraphics[width=0.9\hsize]{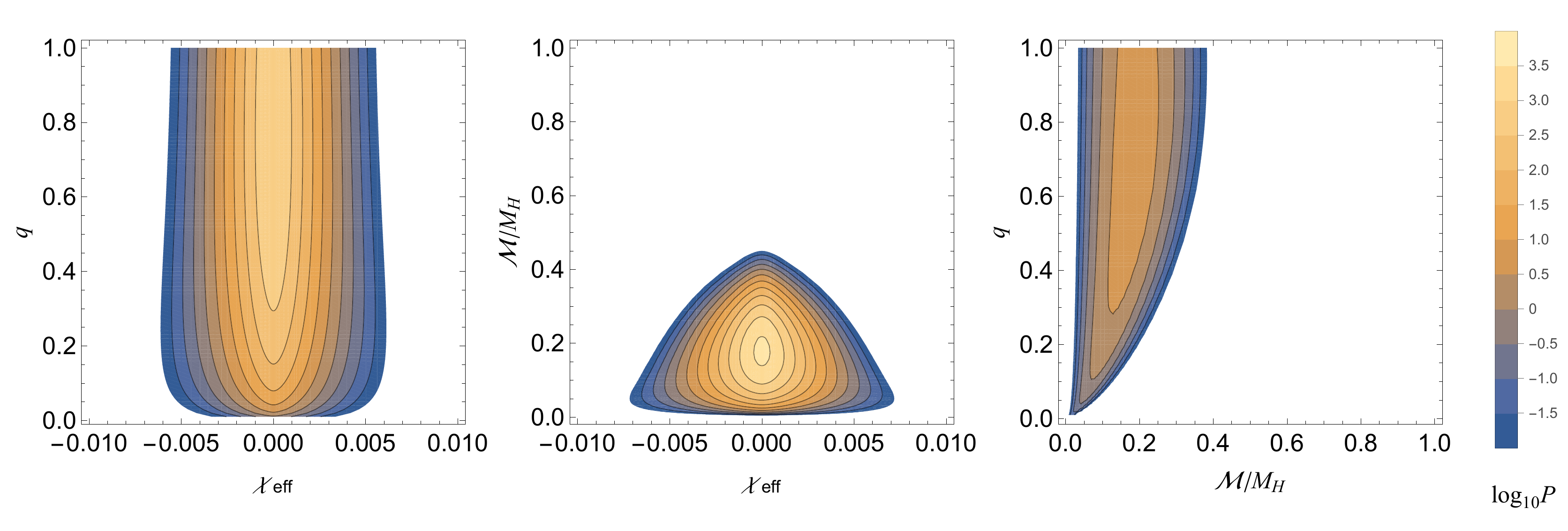}
    \caption{
    Contour plots %Plot 
    of $\log_{10}P(\chi_{\mr{eff}},q)$ (left), $\log_{10}P(\chi_{\mr{eff}},\calM)$ (middle), and $\log_{10}P(\calM,q)$ (right) for $\nu_\uth=10$ and $\sigma=0.192$ which correspond to $f_\PBH\sim0.1\%$ for $M_H\sim M_\odot$. %, and $M_{k_0}=1$.
    }
    \label{fig:PcalMqchi-sgm0192}
\end{figure*}

By integrating it over one of the three variables, one can further obtain the two-variable probabilities $P(\chi_{\mr{eff}},q)$, $P(\calM,\chi_{\mr{eff}})$, and $P(\calM,q)$.
The numerical results are shown in Fig.~\ref{fig:PcalMqchi-sgm0192}.
We take the parameters as $\nu_\uth=10$ and $\sigma_0=0.192$ which correspond to $f_\PBH\sim0.1\%$ for $M_H\sim M_\odot$ \acp{PBH} as discussed in Sec.~\ref{sec:PBH-abundance}.

One can see that the effective spin is distributed in a very narrow region, $|\chi_\mr{eff}|\lesssim 10^{-3}$.
The root mean square is given as $\sqrt{\langle\chi_{\mr{eff}}^2\rangle}=8.41\times10^{-4}$.
This would be because, although the effect of the critical phenomena allows each PBH to spin rapidly so that $a\sim 1$ if the mass is very small, the probability of such small mass is very low as can be seen from Fig.~\ref{fig:PaM-sgm0192}.
The mass ratio is broadly distributed as $0.1\lesssim q\le1$, and the chirp mass has a width $0.1\lesssim{\cal M}/M_{H}\lesssim0.3$ due to the critical behavior~\eqref{eq:critical-behavior} even though we assume an almost monochromatic power spectrum (i.e., a single value for $M_H$).
In addition, in the plot of $P(\chi_{\mr{eff}},\mathcal{M})$, we can see an anti-correlated behavior between  $\sqrt{\langle\chi_\mr{eff}^2\rangle}$
%$\chi_{\mr{eff}}$ 
and $\mathcal{M}$. That is, for smaller $\mathcal{M}$, %\Mag{the distribution of $\chi_{\mr{eff}}$ becomes broader}. 
the root mean square of the effective spin, $\sqrt{\langle\chi_\mr{eff}^2\rangle}(\calM)$, becomes larger.
Actually, by numerical calculation, one can confirm that $\sqrt{\langle\chi_\mr{eff}^2\rangle}(\calM)$ is monotonically decreasing with $\calM$ in the range of $3\times 10^{-4}\lesssim\sqrt{\langle\chi_\mr{eff}^2\rangle}(\calM)\lesssim 3\times 10^{-3}$. This is a result expected from the anti-correlation between $\langle a\rangle$ and $M$ for the single PBH distribution.
%Although the critical phenomena can lead to formation of rapidly spinning PBHs \Red{if their masses are very small as shown in Fig.~\ref{fig:PaM-sgm0192}}, the effective inspiral spin of a PBH binary is typically very small.
%\Red{This is because, even if the secondary PBH has a very small mass and therefore has a Kerr parameter of order unity, $a_2\sim1$, its contribution to $\chi_{\mr{eff}}$ of the binary is suppressed by the very small mass ratio, $q$, according to Eq.~\eqref{eq:chi-q}.}
On the other hand, we find that there is almost no correlation between $\sqrt{\langle\chi_\mr{eff}^2\rangle}\mathblue{(q)}$ and $q$.
In particular, $\sqrt{\langle\chi_\mr{eff}^2\rangle}(q)$ cannot be large even for smaller $q$.
This would be because, even if the secondary PBH has a very small mass and therefore has a Kerr parameter of order unity, $a_2\sim1$, its contribution to $\chi_{\mr{eff}}$ is suppressed by the very small mass ratio, $q$, according to the definition~\eqref{eq:chi-q}.
%To predict \Red{the anti-correlation between $\chi_\mr{eff}$ and $q$} suggested in \citet{Callister_2021}, some effects of binary formation should be incorporated in the current model.

In this paper, as discussed in Sec.~\ref{sec:PBH-abundance}, we have adopted the threshold value of the compaction function, $\calC_{\um,\uth}\simeq2/5$, which leads to $\sigma_0\simeq0.192$ under the assumption $\nu_\uth=10$.
However, this threshold value would include some uncertainty because the value of $\calC_{\um,\uth}$ slightly depends on the initial profile of the perturbation unlike the averaged one~\citep{Escriva_2020b} and the non-zero angular momentum of the collapsing fraction of the universe should make the threshold value higher due to the centrifugal force against the gravitational contraction.
We see how this uncertainty affects the resulting distribution in Appendix~\ref{sec:different-sigma} by taking different values of $\sigma_0$ with the fixed value of $\nu_\uth=10$, which we have determined by using Eq.~\eqref{eq: fPBH}.
We can see that the modification of $\sigma_0$ somehow, but not greatly, changes the widths of the distribution.

%%%sec
\section{Conclusions}
\label{sec:conclusion}

%We have proposed a simplest model of a PBH binary distribution and derived the probability distribution of the parameters, $\chi_{\mr{eff}}$, $q$, and $\calM$, while taking the effect of the critical phenomena of gravitational collapse into account.
In this paper, we formulated the probability distribution of the characteristics (the effective inspiral spin $\chi_\eff$, the mass ratio $q$, and the chirp mass $\calM$ in particular) of \ac{PBH} binaries, taking account of the critical phenomena of gravitational collapse.
First we have derived the distribution of the spin $a$ and the mass $M$ of single PBHs. 
It is basically featured by the scaling relation $a\propto (M/M_H)^{-1/3}$ and PBHs with $M\lesssim 10^{-8}M_H$ can have spins of order unity, while the spin is rather suppressed for $M\gtrsim M_H$.
Under the assumption that two \acp{PBH} of a binary are randomly chosen as a leading order approximation, this single \ac{PBH} distribution is straightforwardly followed by the binary \ac{PBH} distribution.
The resultant probability in $\chi_\eff$, $q$, and $\calM$ is shown in Fig.~\ref{fig:PcalMqchi-sgm0192}.

The first observation is the symmetry under $\chi_\eff\leftrightarrow-\chi_\eff$, which is a direct consequence of our random-choice assumption.
Because of the isotropic distribution of the spin of each PBH, the probability for realizing a binary configuration with the orbital angular momentum ${\bf L}$ should be the same as that with $-{\bf L}$.
This symmetry appears in $\chi_{\mr{eff}}$, Eq.~\eqref{eq:chi-q}, in terms of the polar angle of each spin, $\theta_i$.
That is, the reflection ${\bf L}%\to
\leftrightarrow-{\bf L}$ corresponds to $\theta_i %\to
\leftrightarrow\pi-\theta_i$ leading to the symmetry for $\chi_{\mr{eff}} %\to
\leftrightarrow-\chi_{\mr{eff}}$.
Actually, Eq.~\eqref{eq:P-lncaM-q-chi} depends on $\chi_{\mr{eff}}$ only through the even function of $\chi_{\mr{eff}}$, $T(a_1,a_2,\chi_{\mr{eff}},q)$.
This symmetry implies that at least a certain fraction of black hole binaries have negative values of $\chi_{\mr{eff}}$. 
The negative values of $\chi_{\mr{eff}}$ for black hole binaries with $q\sim 1$ have been indicated 
by the analyses in Refs.~\citep{Callister_2021,theligoscientificcollaboration2021population} although there are only two candidate events~(GW191109\_010717 and GW200225\_060421) with significant support~\citep{theligoscientificcollaboration2021population}. 
The PBH binary scenario would have a potential to explain those negative values.
%To explain the tendency \Red{that a positive value of $\chi_{\mr{eff}}$ might be favored for small $q$}
%of anti-correlation between $\chi_{\mr{eff}}$ and $q$ 
%implied by \citet{Callister_2021} for the O3 data, it would be necessary to take into account some mechanisms of PBH binary formation, such as nonlinear interaction between the orbital angular momentum and the spins.%[ref?]

However, 
the amplitude $\abs{\chi_\eff}$ is found to be very small as $\abs{\chi_\eff}\lesssim8.41\times10^{-4}$, compared to the observed ones $\abs{\chi_\eff}\sim0.1$.
%Although the magnitude of spins of PBHs formed through the critical phenomena can be of order unity, we have found that the effective spins of PBH binaries are distributed in the very \Red{narrow} range $|\chi_{\mr{eff}}|\lesssim0.0015$.
Furthermore, contrary to the anti-correlation between $\abs{\chi_\eff}$ and $\calM$ as expected from the anti-correlation between $a$ and $M$ in the single \ac{PBH} distribution, we found almost no correlation between $\abs{\chi_\eff}$ and $q$.
This would be because, even though the spin of the secondary \ac{PBH} $a_2$ can be large enough if PBH2 is very light and the mass ratio $q$ is very small, the contribution of $a_2$ to $\chi_\eff$ is suppressed by the factor $q$ as can be seen its definition~\eqref{eq:chi-q}.
%\Red{This range is almost independent of $q$ implying that $\chi_{\mr{eff}}$ and $q$ is not correlated. Even for much smaller $q$, which implies the spin of the secondary PBH $a_2$ is expected to be of order unity, $\chi_{\mr{eff}}$ cannot be larger. This would be because the contribution of $a_2$ to $\chi_{\mr{eff}}$ is suppressed by $q$.}\YK{Is there better explanation?}
Therefore, it would be difficult to realize the observed anti-correlation between $\chi_\eff$ and $q$ in our scenario.

As a further consideration, one may include the clustering effect on the \ac{PBH} spatial distribution to alter the random-choice assumption.
%Other effects on binary formation would be also interesting for the realization of a broader distribution in $\chi_{\mr{eff}}$.
For example, it is known that PBHs are clustered when %there is non-Gaussianity of 
the source perturbations are non-Gaussian (see, e.g., \citet{Sasaki:2018dmp}).
Primordial non-Gaussianities may also change the peak statistics and then the spin distribution.
%If such clustering leads to alignment of spins of PBHs, the effective spin of binaries in the cluster would become larger.%[ref?]
Spin evolution through accretion process is also interesting.
%Actually, 
\citet{Luca_2020} shows the evolution can be significant for massive \acp{PBH} $\gtrsim\calO(10)M_\odot$.
The change of pressure of the background fluid is another possibility to enhance the \ac{PBH} spins. 
The pressure $p$ can be reduced from the radiational one $\rho/3$, where $\rho$ is the energy density, during, e.g., the \ac{QCD} phase transition or the possible matter-dominated era in the early universe.
The reduction of pressure allows a non-spherical collapse and the resultant \ac{PBH} can have a large spin~\citep{Harada:2017fjm}.
The \ac{QCD} phase corresponds to $\sim M_\odot$ \acp{PBH} and thus it would have a remarkable relation to merger \ac{GW} events.
%\YT{anything else?}
We leave all these possibilities for future works.

\begin{acknowledgments}
The authors are grateful to 
A. Escriv\`{a}, S. Hirano, T. Kokubu, S. Kuroyanagi, A. Kusenko, and M. Sasaki for their fruitful discussions.
This work is supported by JSPS KAKENHI Grants 
No. JP21K20367 (Y.K.), JP19K03876 (T.H.), JP19H01895 (Y.K., C.Y., T.H.), JP20H05853 (Y.K., C.Y., T.H.), JP20H05850 (C.Y.), JP21K13918 (Y.T.), JP20H01932 (S.Y.) and JP20K03968 (S.Y.)
from the Japan Society for the Promotion of Science.
\end{acknowledgments}

%%%appendix
\appendix
%%%sec
\section{Dimensionless spin parameter \lowercase{$h$}}
\label{sec:def-h}

We have quoted the result for the spin distribution of a single PBH obtained in \citet{Harada_2021}.
In this section, we briefly introduce the relevant quantities for the derivation, especially, the dimensionless spin parameter $h$.
The parameter $h$ was first introduced by %Heavens and Peacock~
\citet{Heavens_1988} and applied to  derivation of PBH spin distribution by %De Luca et al.~
\citet{Luca_2019}.

Let us consider the $3+1$ decomposition of the spacetime,
\bme{
    \dd{s^2}=-\alpha^2(\eta,\bfx)\dd{\eta^2} \\ 
    +\mathfrak{a}^2(\eta)\gamma_{ij}\qty(\dd{x^i}+\beta^i(\eta,\bfx)\dd{\eta})\qty(\dd{x^j}+\beta^j(\eta,\bfx)\dd{\eta}),
}
with a background flat FLRW metric,
\bae{
    \dd{s^2}=\mathfrak{a}^2(\eta)(-\dd{\eta^2}+\dd{x^2}+\dd{y^2}+\dd{z^2}).
}
$\mathmag{\mathfrak{a}}(\eta)$, $\alpha(\eta,\bfx)$, and $\beta^i(\eta,\bfx)$ denote the global scale factor, the lapse function, and the shift vector, respectively.
We assume the matter field to be a single perfect fluid,
\bae{
T^{ab}=\rho u^au^b+p(g^{ab}+u^au^b).
}
On the background spacetime, there are rotational Killing vectors $\phi_i^a=\epsilon_{ijk}(x-x_\mr{pk})^j\delta^{kl}(\partial/\partial x^l)^a$ $(i=1,2,3)$ tangent to a spacelike hypersurface $\eta=\text{const}$.
For a region $\Sigma$ on the spacelike hypersurface, the conserved angular momentum $S_i(\Sigma)$ of the matter contained in $\Sigma$ can be defined as
\bae{
    S_i(\Sigma)\coloneqq{}&\frac{1}{16\pi G}\int_{\partial\Sigma}\epsilon_{abcd}\nabla^c(\phi_i)^d \nonumber \\
    ={}&-\frac{1}{8\pi G}\int_\Sigma R^{ab}n_a(\phi_i)_b\dd{\Sigma} \nonumber \\
    ={}&-\int_\Sigma T^{ab}n_a(\phi_i)_b\dd{\Sigma},
}
where the Einstein equation is used in the last equality.
For primordial black hole formation, we suppose $\Sigma$ to be a region that will collapse into a black hole, and the black hole mass and angular momentum are estimated as those of matter in $\Sigma$.
Here we assume that the region $\Sigma$ is given by
\bae{
\Sigma= \left\{\bfx\mid\delta(\bfx)>f\delta_\mr{pk}
\right\},
}
with some positive constant $f$ less than unity.

Around the peak, the density contrast, which we assume to be a Gaussian random field, is expanded as
\bae{
\delta\simeq \delta_\mr{pk}+\frac{1}{2}\zeta_{ij}(x-x_\mr{pk})^i(x-x_\mr{pk})^j,
}
where 
\bae{
    \zeta_{ij}\coloneqq\left.\frac{\partial^2\delta}{\partial x^i\partial x^j}\right|_{\bfx=\bfx_\mr{pk}}.
}
Taking $x$, $y$, and $z$ axes as the principal directions of $\zeta_{ij}$, we have
\bae{
\delta\simeq\delta_\mr{pk}-\frac{1}{2}\sigma_2\sum_{i=1}^3\lambda_i((x-x_\mr{pk})^i)^2,
}
where $\sigma_j$ is defined below Eq.~\eqref{eq:Ph-Heavens} and $\lambda_1\leq\lambda_2\leq\lambda_3$ are the eigenvalues of $-\zeta_{ij}/\sigma_2$.
As a result, $\Sigma$ is given as an ellipsoid with the three axes,
\bae{
a_i^2=2\frac{\sigma_0}{\sigma_2}\frac{1-f}{\lambda_i}\nu.
}

Expanding the fluid 3-velocity $v^i\coloneqq u^i/u^0$ as
\bae{
    v^i-v_\mr{pk}^i\simeq v^i{}_j(x-x_\mr{pk})^j,
}
we obtain
\bae{
    S_i(\Sigma)
    &\simeq(1+w)\mathfrak{a}^4\rho_b\epsilon_{ijk}v^k{}_l\int_\Sigma(x-x_\mr{pk})^j(x-x_\mr{pk})^l\dd[3]{x}\nonumber\\
    &=(1+w)\mathfrak{a}^4\rho_b\epsilon_{ijk}v^k{}_lJ^{jl},
}
where $w=p/\rho$ and
\beae{
    v^k{}_l\coloneqq{}&\left.\frac{\partial v^k}{\partial x^l}\right|_{\bfx=\bfx_\mr{pk}}, \\
    J^{jl}\coloneqq{}&\int_\Sigma(x-x_\mr{pk})^j(x-x_\mr{pk})^l\dd[3]{x} \\
    ={}&\frac{4\pi}{15}a_1a_2a_3\,\diag(a_1^2,a_2^2,a_3^2).
}

For PBH formation, we focus on a growing mode of the perturbation.
The time dependence of the perturbation is investigated in \citet{Harada_2021}.
According to it, the average of the spin magnitude is decomposed as
\bae{
\sqrt{\langle S_iS^i\rangle}=S_\mr{ref}\sqrt{\langle s_e^is_{ei}\rangle},
}
where 
\beae{
\label{eq:AppA-Sref}
    S_\mr{ref}(\eta)={}&(1+w)\mathfrak{a}^4\rho_bg(\eta)(1-f)^{5/2}R_*^5,\\
    \vec{s}_{e}={}&\frac{16\sqrt{2}\pi}{135\sqrt{3}}\left(\frac{\nu}{\gamma}\right)^{5/2}\frac{1}{\sqrt{\Lambda}}(-\alpha_1\tilde{v}_{23},\alpha_2\tilde{v}_{13},-\alpha_3\tilde{v}_{12}), \\
    \alpha_1={}&\frac{1}{\lambda_3}-\frac{1}{\lambda_2} \qc
    \alpha_2=\frac{1}{\lambda_3}-\frac{1}{\lambda_1} \qc
    \alpha_3=\frac{1}{\lambda_2}-\frac{1}{\lambda_1}, \\
    \Lambda\coloneqq{}&\lambda_1\lambda_2\lambda_3 \qc
    R_*\coloneqq\sqrt{3}\frac{\sigma_1}{\sigma_2}.
}
The function $g(\eta)$ defined by
\bae{
    \Braket{(v^k{}_l(\eta))^2}=g^2(\eta)\Braket{(\tilde{v}^k{}_l)^2},
}
represents the time evolution of the velocity field for every $k,l$, where the time-independent variable $\tilde{v}^k{}_l$ is defined by
\bae{
    \tilde{v}^i{}_j\coloneqq-\frac{1}{\sigma_0}\int\frac{\dd[3]{%\bf 
    k}}{(2\pi)^3}\frac{k^ik_j}{k^2}\delta_{\bf k}\ee^{i{\bf k\cdot %r
    \mathblue{\bfz}}},
}

For large $\nu$ limit, the dimensionless spin parameter $h$ is defined by
\bae{
\label{eq:def-h}
    s_e\coloneqq\sqrt{\vec{s}_e\cdot\vec{s}_e}=\frac{2^{9/2}\pi}{5\gamma^6\nu}\sqrt{1-\gamma^2}h.
}
$h$ is useful to investigate the probability distribution of the spin.
%Heavens and Peacock~
\citet{Heavens_1988} numerically derived the probability distribution as
\bae{
    &P_h(h)\dd{h}=563h^2 \nonumber \\
    &\times\exp\bqty{-12h+2.5h^{1.5}+8-3.2(1500+h^{16})^{1/8}}\dd{h}.
}
Recently, %De Luca et al.~
\citet{Luca_2019} gave another fitting formula which agrees with the above one very well.
In this paper we adopt the former one because of the regular behavior in the limit $h\to0$.

The total angular momentum $S_i(\Sigma)$ will become that of the resulting PBH after the formation.
The region $\Sigma$, which will collapse into a PBH, would be specified when the contraction of matter is decoupled from the expansion of the universe.
This is called turn around.
We denote the time of turn around as $\eta_\mr{ta}$.
Defining the dimensionless reference spin value at turn around as
\bae{
\label{eq:AppA-Aref}
A_\mr{ref}(\eta_\mr{ta})=\frac{S_\mr{ref}(\eta_\mr{ta})}{GM_\mr{ta}^2},
}
where $M_\mr{ta}$ is the mass inside $\Sigma$ at the turn around, we can estimate the initial dimensionless Kerr parameter $a$ of the resulting PBH.
For the radiation domination, %Ref.~
\citet{Harada_2021} found, in Eq.~(22) of it, the simple expression of $A_\mr{ref}(\eta_\mr{ta})$ as
\bae{
	A_\mr{ref}(\eta_\mr{ta})\simeq\frac{1}{24\sqrt{3}\pi}x_\mr{ta}^2(1-f)^{-1/2}\abs{{T_v}_{\mr{CN}}(k_0,\eta_\mr{ta})}\sigma_H,
}
where $x_\mr{ta}=k_0\eta_\mr{ta}$, 
%$f$ is the fraction of the mass inside the horizon which collapses into a PBH, 
${T_v}_{\mr{CN}}(k_0,\eta)$ is the transfer function of the mode of the velocity field with wavelength $k_0$ in the conformal Newtonian gauge, and $\sigma_H$ denotes $\sigma_0$ when the initial time of the evolution of the cosmological perturbation is set to the horizon entry.
For radiation domination, the factors were numerically estimated as $x_\mr{ta}\simeq2.14$ and ${T_v}_{\mr{CN}}(k_0,\eta_\mr{ta})\simeq0.622$ in Sec.~3.3 of %Ref.~
\citet{Harada_2021}.
Using the relation between $M_\mr{ta}$ and $f$ derived in %Ref.~
\citet{Harada_2021},
\bae{
\label{eq:AppA-Mta-f}
	M_\mr{ta}\simeq\frac{\sqrt{6}}{x_\mr{ta}}(1-f)^{3/2}M_H,
}
and identifying $M_\mr{ta}$ with $M$ and $\sigma_H$ with $\sigma_0$, we have
\bae{
	A_\mr{ref}(\eta_\mr{ta})\simeq 2.28\times10^{-2}\sigma_0\left(\frac{M}{M_H}\right)^{-1/3}.
}
Then, we obtain the Kerr parameter of a PBH by putting $a=\sqrt{S_iS^i}/GM^2=A_\mr{ref}s_e=Ch$ in terms of $h$ with a coefficient $C$, where
\bae{
    C&=\frac{2^{9/2}\pi}{5\gamma^6\nu}\sqrt{1-\gamma^2}A_\mr{ref}(\eta_\mr{ta})\nonumber\\
    &=3.25\times 10^{-2}\sqrt{1-\gamma^2} \,\sigma_0\left(\frac{M}{M_H}\right)^{-1/3}\left(\frac{\nu}{10}\right)^{-1}.
}
We have set $\gamma^6\simeq1$ in the last equality.

%%%sec
\section{Correction due to peak finding condition}
\label{sec:peak-number}

We regard the density contrast $\delta$ as a Gaussian random field.
This implies that its scaling $\delta/\sigma_0$ is also a Gaussian random field.
However, if we focus on its peaks, the probability distribution of the peak values, $\nu=\delta_\mr{pk}/\sigma_0$, are not given by a Gaussian function because it is corrected by {\it the peak-finding condition}.

According to %Ref.~
\citet{Bardeen:1985tr}, the number density of the peaks with the value in $(\nu,\nu+\dd{\nu})$ is given by
\bme{
    \mathcal{N}_\mr{pk}(\nu)\dd{\nu}=\frac{1}{(2\pi)^2}\left(\frac{\sigma_2}{\sqrt{3}\sigma_1}\right)^3
    \ee^{-\nu^2/2} \\ 
    \times\left[\int^\infty_0\dd{x}f(x)\frac{\exp[-(x-\gamma\nu)^2/2(1-\gamma^2)]}{[2\pi(1-\gamma^2)]^{1/2}}\right]\dd{\nu},
}
where $x\coloneqq-\nabla^2\delta/\sigma_2$ is the width of the peak and $f(x)$ is a function behaving $f(x)\to x^3-3x$ for large $x$.
Note that $x$ is also a statistical variable which is, in general, independent of $\nu$.

For the perfect correlation of $\nu$ and $x$, $\gamma\to1$, it reduces to
\bae{
    \mathcal{N}_\mr{pk}(\nu)\dd{\nu}=\frac{1}{(2\pi)^2}\left(\frac{\sigma_2}{\sqrt{3}\sigma_1}\right)^3
    \ee^{-\nu^2/2}f(\nu)\dd{\nu}.
}
In a finite volume $V$, the number of peaks that will collapse into PBHs, i.e., peaks of $\nu>\nu_\uth$, is given by
\bae{
    N_\mr{PBH}
    =V\int^\infty_{\nu_\uth}\mathcal{N}_\mr{pk}(\nu)\dd{\nu}.
}
The number of peaks in the range $(\nu,\nu+\dd{\nu})$ in $V$ is given by
\bae{
    n_\mr{PBH}(\nu)\dd{\nu}
    =V\mathcal{N}_\mr{pk}(\nu)\dd{\nu}.
}
Then, in the volume $V$, the probability distribution for one to find a peak in the range $(\nu,\nu+\dd{\nu})$ from all the peaks greater than the threshold is given by
\bae{
    \label{eq:Pnu-fcorrection}
    P_\nu(\nu)\dd{\nu}
    =\frac{n_\mr{PBH}(\nu)\dd{\nu}}{N_\mr{PBH}}
    =\frac{\ee^{-\nu^2/2}f(\nu)\dd{\nu}}{\int^\infty_{\nu_\uth}\ee^{-\bar{\nu}^2/2}f(\bar{\nu})\dd{\bar{\nu}}}.
}
Therefore, the peak finding-condition is given by a Gaussian function with the correction factor $f(\nu)$.
However, for peaks that will collapse into PBHs, the values of $\nu$ are always large such that $\nu>\nu_\uth\sim10$.
Thus, the Gaussian factor $\ee^{-\nu^2/2}$ rapidly decays for larger $\nu$ in the range $(\nu_\uth,\infty)$ and  contributes to the probability distribution of PBH binaries, Eq.~\eqref{eq:P-lncaM-q-chi}, only if $\nu\sim\nu_\uth$.
Then, we can regard the factor $f(\nu)\sim f(\nu_\uth)$ as a constant, which contributes to the overall factor.
As a conclusion, we approximate the probability distribution as
\bae{
    \label{eq:Pnu-approx}
    P_\nu(\nu)\dd{\nu}
    =\frac{\ee^{-\nu^2/2}f(\nu)\dd{\nu}}{\int^\infty_{\nu_\uth}\ee^{-\bar{\nu}^2/2}f(\bar{\nu})\dd{\bar{\nu}}}
    \simeq\frac{\ee^{-\nu^2/2}\dd{\nu}}{\int^\infty_{\nu_\uth}\ee^{-\bar{\nu}^2/2}\dd{\bar{\nu}}},
}
as in Eq.~\eqref{eq: Pnu for PBH}.

As discussed in the end of Sec~\ref{sec:distribution-binary}, the current PBH binary model has an uncertainty because, for example, we have applied numerical results obtained in works which assume spherical symmetry.
We estimate the effect of this uncertainty in Appendix~\ref{sec:different-sigma} and find that it somehow, but not greatly, changes the widths of the distribution.
Thus, the above approximation, Eq.~\eqref{eq:Pnu-approx}, would not matter compared with this uncertainty.

%%%sec
\section{Modification of $\sigma_0$}
\label{sec:different-sigma}

\begin{figure*}
    \centering
    \includegraphics[width=%450pt
    0.9\hsize]{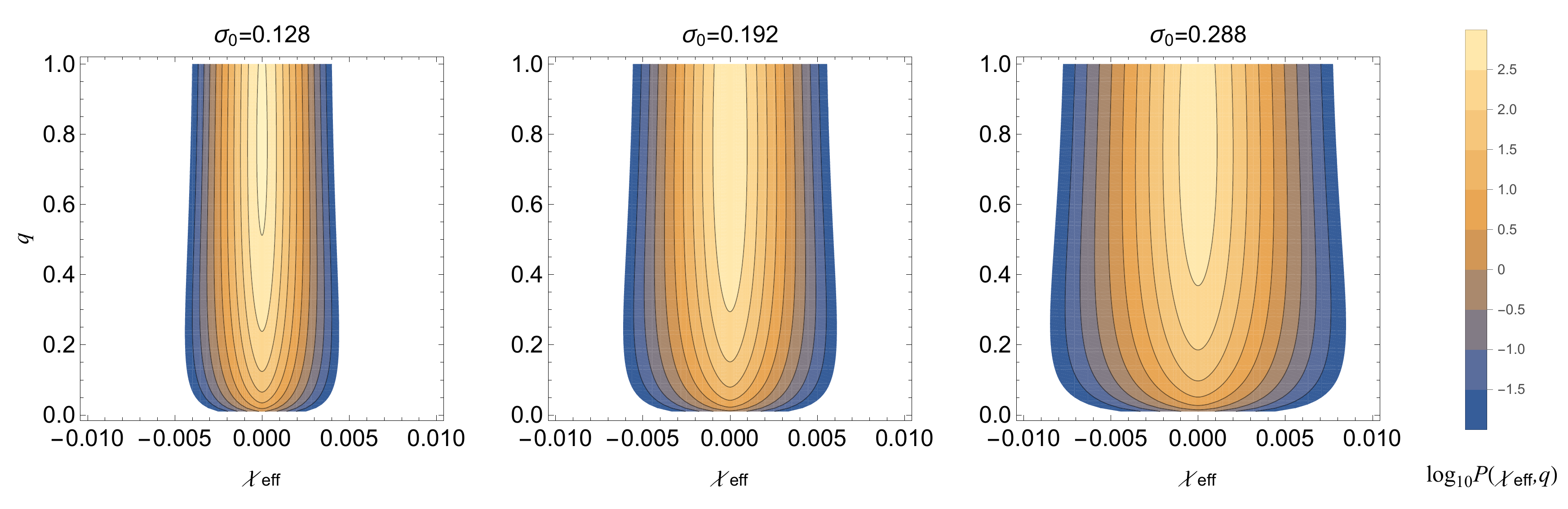}
    \caption{Contour plots of $\log_{10}P(\chi_{\mr{eff}},q)$ for $\sigma_0=0.128$, $0.192$, and $0.288$.
    }
    \label{fig:PchiqAll}
\end{figure*}

\begin{figure*}
    \centering
    \includegraphics[width=%450pt
    0.9\hsize]{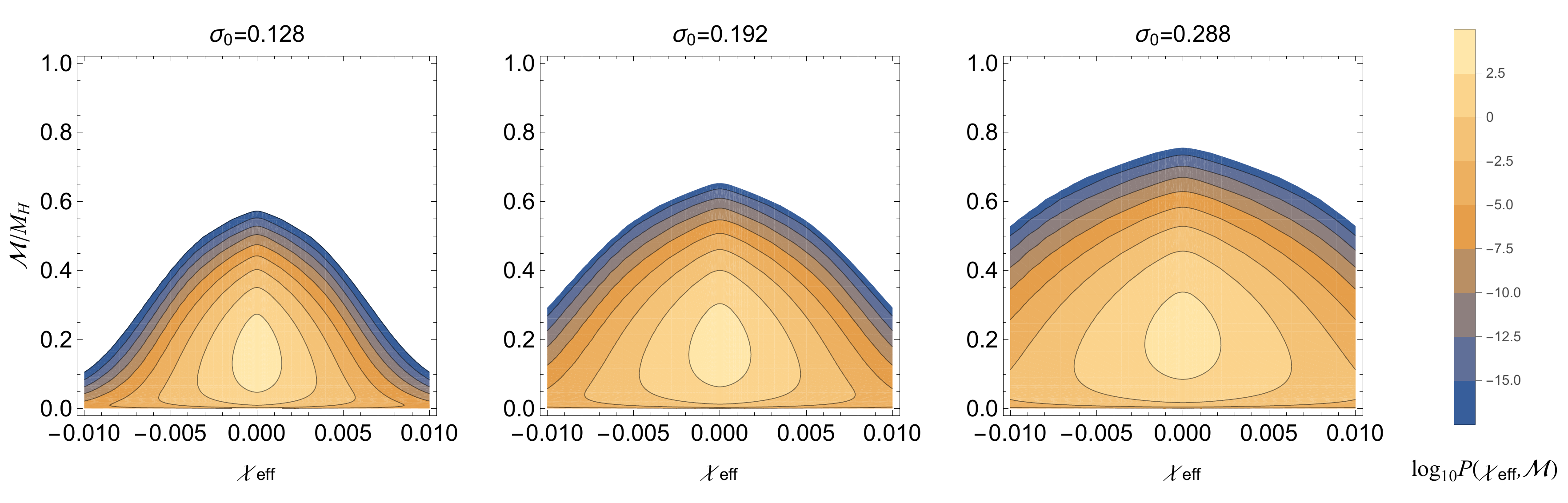}
    \caption{Contour plots of $\log_{10}P(\chi_{\mr{eff}},\calM)$ for $\sigma_0=0.128$, $0.192$, and $0.288$.
    }
    \label{fig:PchimAll}
\end{figure*}

\begin{figure*}
    \centering
    \includegraphics[width=%450pt
    0.9\hsize]{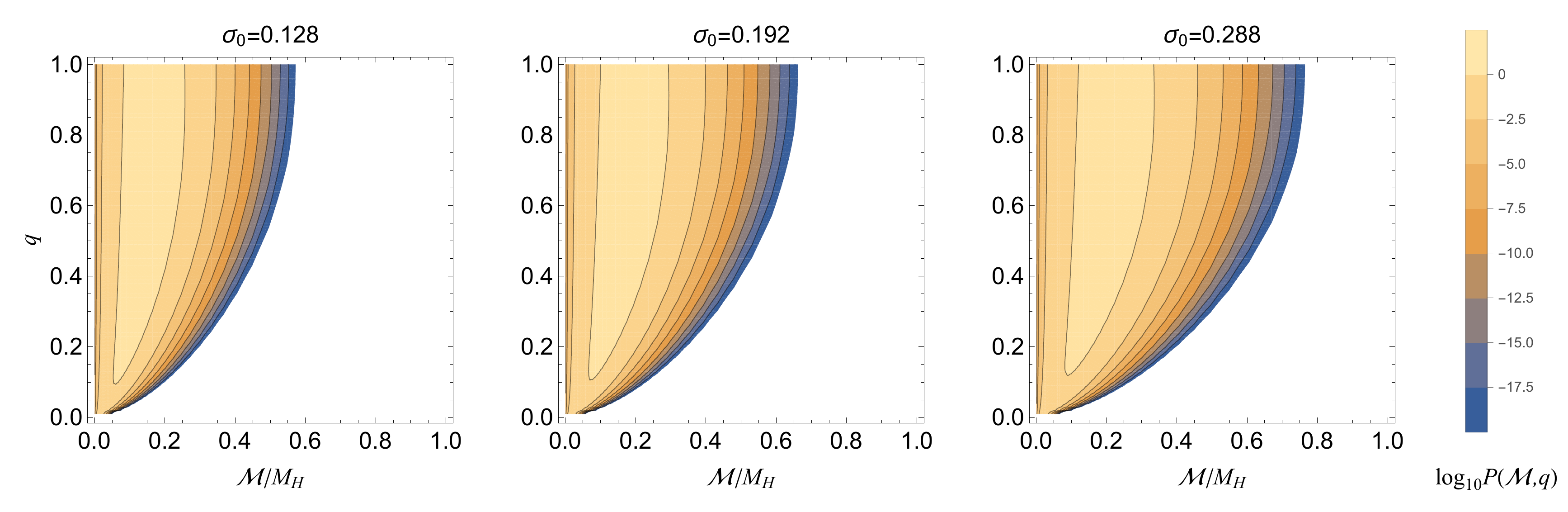}
    \caption{Contour plots of $\log_{10}P(\calM,q)$ for $\sigma_0=0.128$, $0.192$, and $0.288$.
    }
    \label{fig:PmqAll}
\end{figure*}

As a modification of the peak threshold due to the uncertainty of the value of $\calC_{\um,\uth}$, we here show the numerical results of $P(\calM,q,\chi_{\mr{eff}})$ with different values of $\sigma_0$.
We take the values as $\sigma_0=0.128$ ($=(2/3)\times0.192$), $0.192$ (the value taken in the main part), and $0.288$ ($=(3/2)\times0.192$). %\YT{what's the motivation of these values?}
The results in Figs.~\ref{fig:PchiqAll}--\ref{fig:PmqAll} show that the distribution is somehow broadened (narrowed) for larger (smaller) $\sigma_0$ with fixed $\nu_\uth=10$.
In particular, the widths in $\chi_{\mr{eff}}$ of $P(\chi_{\mr{eff}},q)$ are about $0.001$, $0.0015$, and $0.002$ for $\sigma_0=0.288$, $0.192$, and $0.128$, respectively.
Thus, we conclude that the uncertainty in the value of $\calC_{\um,\uth}$ does not greatly affect the distribution.

%\bibliography{sample}
%\bibliographystyle{/usr/local/texlive/2018/texmf-dist/pbibtex/bst/JHEP}
%
\bibliography{draft}
\bibliographystyle{aasjournal}

\end{document}